***Short title for running head:*** Erndwein et al. – Field-based Mechanical Phenotyping

**Field-based mechanical phenotyping of cereal crops to assess lodging resistance[1]**

Lindsay Erndwein[2], Douglas D. Cook[3], Daniel J. Robertson[4], and Erin E. Sparks[2,5]

[2] Department of Plant and Soil Sciences and the Delaware Biotechnology Institute, University of Delaware, Newark, DE 19711 USA

[3] Department of Mechanical Engineering, Brigham Young University, Provo, UT 84602 USA

[4] Department of Mechanical Engineering, University of Idaho, Moscow, ID, 83844 USA

Email address:          LE:  erndwein@udel.edu

                        DDC:  d.cook@byu.edu

                        DJR:  danieljr@uidaho.edu

                        EES:  esparks@udel.edu

[5] Author for correspondence:  esparks@udel.edu

# ABSTRACT

Plant mechanical failure, also known as lodging, is the cause of significant and unpredictable yield losses in cereal crops. Lodging occurs in two distinct failure modes - stalk lodging and root lodging. Despite the prevalence and detrimental impact of lodging on crop yields, there is little consensus on how to phenotype plants in the field for lodging resistance and thus breed for mechanically resilient plants. This review provides an overview of field-based mechanical testing approaches to assess stalk and root lodging resistance. These approaches are placed in the context of future perspectives. Best practices and recommendations for acquiring field-based mechanical phenotypes of plants are also presented.

**Key words:**   Anchorage; Bending; Rind Penetration; Lodging; Root; Stalk

# INTRODUCTION

Cereal crops are faced with a complex mechanical challenge - they must be rigid enough to support their own weight, but flexible enough to be resilient in the face of external forces (e.g. wind) (Gardiner et al., 2016). Mechanical failure of cereal crops is known as lodging and refers to the "permanent displacement of plants from their vertical stance" (Rajkumara, 2008). Lodging is multifactorial and stochastic, but has been reported to cause up to 80% yield losses, depending on the crop and field location (Rajkumara, 2008; Berry et al., 2004), and even if harvested, lodging can reduce grain quality (Mizuno et al., 2018). Factors underlying plant susceptibility to lodging include meteorological factors (e.g. wind, rain, and hail), field management practices (Rajkumara, 2008), plant architecture (Stamp and Kiel, 1992; Brune et al., 2018), and plant biomechanics (Robertson et al., 2016). Herein we provide a review of current field-based mechanical testing approaches used to assess lodging resistance in cereal crops.

There are two types of lodging that are distinguished by the point of mechanical failure - stalk lodging and root lodging (Berry et al., 2004). In cereal crops, stalk lodging refers to breakage of the stem below the height of the flower and root lodging refers to failure at the root-soil interface. In maize, stalk failure is further distinguished by the point of failure as green (brittle) snap or stalk lodging. Green snap refers to stalk breakage at the stem node prior to flowering, whereas stalk lodging refers to stem internode buckling and occurs at late plant stages. Root lodging can occur at any plant stage, however yield losses become more severe as plants mature or as time to harvest decreases (e.g. Carter and Hudelson, 1988). From an agronomic perspective, stalk and root lodging often occur in the same field and are rarely differentiated by growers.

From a plant breeding/phenotyping perspective, stalk and root lodging are distinct in both their failure types and failure mechanisms (Berry et al., 2003a). As such, the genetic and environmental underpinnings of stalk lodging resistance and root lodging resistance are expected to be distinct. Mechanical phenotyping methods used to assess each type of lodging are likewise distinct. Herein, we review field-based mechanical phenotyping approaches to quantify stalk lodging resistance and root lodging resistance in cereals. Recommendations, best practices and future research directions are highlighted. The reader is referred to other focused review articles for more information on the general topic of lodging (Berry et al., 2004; Rajkumara, 2008; Berry 2013; Khobra et al., 2018), the impact of wind forces on plants (Gardiner et al., 2016), root anchorage (Stubbs et al., 2019), laboratory-based mechanical phenotyping of stalks/stems (Shah et al., 2017) and plant biomechanics (Niklas, 1992; Niklas and Spatz, 2012), which fall outside the scope of this review.

## FAILURE ANALYSIS OF NATURALLY OCCURRING LODGING

Understanding the failure patterns associated with lodging is critical to identify breeding targets for lodging resistance. Lodging can manifest by different failure mechanisms, which can provide insight into the most appropriate mechanical phenotyping strategies (Robertson et al., 2015a). In general, stalk lodging failure patterns can vary spatially (location of failure within the plant) and temporally (across the plants lifespan). For example, small grains tend to buckle at the lower internodes (Mulder, 1954; Laude and Pauli, 1956; Neenan and Spencer-Smith, 1975; Fig. 1A), but in barley and oats, buckling of the middle internodes is common and failure can even occur near the peduncle (White, 1991; Fig. 1B). In contrast, large grains like maize tend to fail near a node, but the specific failure pattern can differ by growth stage. For example, analysis of mid-season maize shows that plants fail at the node in a green or brittle snapping pattern (Elmore et al., 2005; Fig. 1C). Whereas, late-season maize stalk lodging primarily involves a mechanism known as brazier buckling (Robertson et al., 2015a; Fig. 1D). Understanding these different failure patterns has been essential to the development of testing protocols and phenotyping methods that reproduce natural failure types and patterns (Robertson et al., 2014; Robertson et al., 2015a).

Plant anchorage is achieved through interactions of roots and soil. Roots can act as tethers in tension or compression, with those tethers failing during root lodging. In addition, a root ball can be formed (i.e., a cohesive root-soil structure) that rotates out of the soil during plant failure. This is similar to uprooting observed in large trees (Easson et al., 1992). Failure during induced root lodging in winter wheat (Crook and Ennos, 1993, 1994) and maize (Ennos et al., 1993) is characterized by the tensile anchorage model – both crops demonstrate buckling of roots on the leeward side (away from the applied force). A study of root failure during natural lodging in wheat also found a failure of tensile anchorage with roots breaking and also roots pulling free of the soil (Easson et al., 1992; Fig. 2A,B). Our unpublished observations of root failure in maize after lodging are consistent with these results. These results favor a model of tensile anchorage over the model of root ball anchorage. However, soil composition may influence the type of failure and additional studies are needed to fully understand the potential variations of root failure patterns.

Considering the variation in failure modes described above, quantifying lodging resistance in the field is not a simple nor singular task. Several approaches have been developed to evaluate lodging resistance in the field, including artificial wind and devices that measure proxies of stalk or root failure. The following sections review the methods that have been used to assess cereal crops for lodging resistance in the field.

## ARTIFICIAL WIND TO EVALUATE STALK AND ROOT LODGING

Artificial wind sources that attempt to mimic the natural weather patterns associated with plant failure have been used to assess lodging resistance. One early study tested the effectiveness of a mobile wind source (an airplane propeller driven by an automobile engine) to evaluate lodging resistance in wheat, oats, and barley (Harrington and Waywell, 1950). This study found that while the artificial wind experiments provide some value to assess lodging resistance, the large size and low-throughput of the wind source made this an unsatisfactory approach to study lodging. More recent studies have developed wind machines to study lodging in wheat (Sterling et al., 2003), maize (Wen et al., 2019), and rice (Shrestha et al., 2020). In wheat, a wind tunnel was constructed with a portable wooden enclosure and 6 axial fans mounted on a mobile trailer. This setup was determined to provide an accurate simulation of natural wind conditions (Sterling et al., 2003). This wind tunnel revealed interesting differences in the timing of lodging – namely that stalk lodging occurs instantaneously and root lodging occurs progressively (Sterling et al., 2003). However, additional studies with this wind tunnel were not found in the literature. For maize, a mobile wind machine was constructed of a high-speed fan set on parallel rails for mobility and used to evaluate stalk lodging as an outcome of variable wind speeds (Wen et al., 2019). This study showed that the failure wind speed varies based on maize variety, but did not attempt to link these results directly to the incidence of lodging (Wen et al., 2019).

Recently a device called Blaster, that is a combination wind machine and rain simulator (Shrestha et al., 2020) has been developed. The device was applied across twenty rice cultivars and three fields seasons, and shows a high prediction of natural stalk lodging when evaluated for a subset of eight genotypes (Shrestha et al., 2020). The bending moment of the lower internodes, as measured by a prostrate tester (see below for additional description of prostrate testers used to evaluate root lodging resistance) was the best single trait predictor of stalk lodging induced by Blaster. However, a compound trait, named "Lodging Resistance Index" (Bending Moment at the Internode / [Above-ground Fresh Weight * Culm Length]; Ookawa and Ishihara, 1992), was the best predictor of lodging induced by Blaster with an $R^2$ of 65-73% depending on windspeed (Shrestha et al., 2020). This study by Shrestha et al., also represents the first comprehensive analysis of wind-induced stalk lodging compared to natural stalk lodging and provides a solid biological basis for genetic variation in stalk lodging resistance. However, it remains unclear if the addition of water to the wind simulator has a significant impact on lodging or whether wind alone could be used for future evaluations. One study of natural wheat lodging in China reported that the combination of wind and rain was related to a higher percent of lodging than either factor alone (Niu et al., 2016), suggesting that the addition of water to wind simulators maybe critical for their successful application in understanding lodging resistance.

The approaches described above all consist of a wind source that is able to be moved from one location to the next, but is static relative to the field. In other words, there is a single point source of wind that is applied in a gradient across the plants. A major advancement was realized for commercial breeding applications, when Pioneer Hi-Bred (now Corteva) developed a mobile wind machine called 'Boreas' to select for green snap, root lodging, and stalk lodging resistance in maize (Barrerio et al., 2008). The basis of Boreas is a wind generator that can move throughout a field, and can apply varying durations and velocities of wind to simulate the environmental conditions associated with each type of lodging (Barrerio et al., 2008). Using Boreas to simulate thunderstorm conditions was reported as a strong indicator of green snap events (Cooper et al., 2014). However, as Boreas is used in commercial applications and protected by patents, the details of the device and research data obtained from this platform are not readily available.

A major consideration in the construction and cost of an artificial wind system is acquiring the desired wind speeds. Theoretically, the failure wind speed of cereals was calculated as 11.6 m/s at the canopy (Baker, 1995), however, this likely varies dramatically between crop type, within different genotypes of the same crop, under different moisture conditions, and is also dependent on planting density. Failure wind speeds have not been directly calculated for many crops, and thus the minimum needed wind speeds for wind simulators is an open question. In maize the failure wind speeds were calculated between 16 m/s and 30 m/s depending on the variety, but this study relied on purely wind-induced failure without soil saturation (Wen et al., 2019) It is likely that failure wind speeds would be reduced with the addition of soil moisture to these studies.

Reported wind speeds achieved by the artificial wind sources described above are: Sterling et al. (2003) - 8.5 m/s (up to 10 m/s for gusts), Wen et al. (2019) - 30 m/s, Blaster - 16.7 m/s (converted from 60 km/h; Shrestha, et al., 2020), and Boreas - 45 m/s (Barrerio et al., 2008). Thus, the wind tunnel of Sterling et al. (2003) would likely not generate enough wind speed to evaluate lodging in maize, but is suitable for smaller stature crops. Although limited in their application (likely because of the cost and expertise need to construct them), artificial wind approaches can provide valuable information about lodging resistance in cereal crops.

## MECHANICAL METHODS TO EVALUATE STALK LODGING

### *RIND PENETRATION*

The most common approach for assessing stalk lodging resistance is measuring rind penetration resistance. This measurement involves piercing the stalk rind with a probe attached to a digital force gauge (Flint-Garcia et al., 2003a; Flint-Garcia et al., 2003b; Peiffer et al., 2013) and recording the maximum force required to penetrate the rind. This method has been used throughout most of the 20th century and dates back to at least 1935 (Khanna, 1935). However, there are conflicting reports of the utility of the rind penetration procedure to evaluate lodging resistance and it is not widely used by commercial breeding programs. Some studies show that rind penetration resistance is highly correlated with stalk lodging resistance (Anderson and White, 1994; Dudley, 1994), while others show that rind penetration is weakly correlated with stalk lodging resistance (McRostie and MacLachlan, 1942; Butrón et al., 2002; Gou et al., 2007; Hu et al., 2012; Robertson et al., 2017). One of the studies with weak correlation compared the results from rind penetration resistance in maize to laboratory-based stalk 3-point bending strength measurements (Robertson et al., 2017), which closely mirror failure patterns of naturally stalk lodged plants (Robertson et al., 2014; Robertson et al., 2015b). In this analysis, rind penetration resistance accounted for less than 20% of the variation in stalk bending strength (Robertson et al., 2017).

These conflicting results about the utility of rind penetration resistance to predict stalk lodging resistance could be attributed to the fact that rind penetration resistance measurements do not quantify the effect that stalk morphological properties have on stalk lodging resistance. From a biomechanical perspective, the stem diameter, cross-sectional area and rind thickness all influence stalk lodging resistance (Robertson et al., 2017). Several studies investigating the genetic architecture of rind penetration resistance have shown that there is not a direct correlation between rind penetration and other morphological features of importance for stalk lodging resistance (Butrón et al., 2002; Martin et al., 2004; Gibson et al., 2010; Hu et al., 2012; Li et al., 2014; Ma et al., 2014). Further, it has been suggested that the relationship between rind penetration resistance and stalk strength is highly dependent on growing conditions such as planting density, genotype, and location (Robertson et al., 2016). A primary reason for the contrasting reports on the effectiveness of rind penetration resistance is likely due to the lack of published testing standards. For example, the geometry of the penetrating probe and the rate of force application are expected to significantly impact measurements, but these factors are rarely reported in papers that utilize the rind penetration method.

*BENDING TESTS*

In considering the natural failure pattern of lodged stalks, several field-based measures of stalk bending stiffness and stalk bending strength have been developed. The original field-based bending test was achieved by fastening different weighted chains to the base of an oat panicle and measuring the stalk displacement from horizontal (Grafius and Brown, 1954). With this information, the authors calculated a metric of lodging resistance (calculated as the torque a plant can resist / applied torque), which showed a moderate correlation with natural lodging (Grafius and Brown, 1954). These experiments were low-throughput and laborious, thus a series of semi-automated field-based bending devices were subsequently developed. These devices can be divided into two categories based on whether they have been applied to small grains (e.g. wheat, rice, oat, or barley) or large grains (e.g. sorghum or maize).

A primary challenge with field-based mechanical testing of small grains is the inability of a single stem to provide a sufficient amount of resistance to reliably detect with a load sensor. To overcome this limitation, multiple plants are tested together in small grain applications. One device (referred to here as Berry's Device) was developed to study winter wheat lodging and consists of a hand-held force meter with a load cell attached to a push bar to measure the force required to push over multiple plants (Fig. 3A, Table 1; Berry et al., 2003b). The measurement obtained from this device is the force applied to reach a discrete angle. Recently, this device was modified by another group to reduce the weight, and continuously record data as plants are bent (Fig. 3B, Table 1; Jo Heuschele et al., 2019). This updated device, called the Stalker, was created to differentiate between different management practices in wheat and between different small grains (wheat, oat and barley), however the results of these trials are not reported in the manuscript (Jo Heuschele et al., 2019).

For larger grains, several different devices have been developed to test individual plant biomechanics in the field. The first device (referred to here as Guo's Device) was developed to non-destructively measure the forces required to bend maize stalks across a set of discrete angles (Fig.3C, Table 1; Guo et al., 2018; Guo et al., 2019). In this device, a controller module with a strain sensor is connected by a belt to a second unit fixed to the stalk. The controller module is pulled to discrete angles ranging from 0- to 45-degrees and the maximum equivalent force recorded. This force was shown to have a strong negative correlation with the incidence of stalk lodging in maize (Guo et al., 2018; Guo et al., 2019).

A second device, DARLING (Device for Assessing Resistance to Lodging in Grains), was also developed to assess stalk biomechanics in larger cereal crops (Cook et al., 2019). This device (DARLING, Fig. 3D, Table 1) collects continuous force-rotation data and consists of a vertical support with a control box mounted at the top, a horizontal footplate attached by a hinge at the base, and an adjustable height load cell. To use, the operator places a stalk in contact with the load cell, and a foot on the hinged base to anchor the device to the ground. Using this device, the stalk can be tested in two modes, either (1) bent until failure to obtain stalk bending strength, or (2) bent within the linear-elastic range of the material to obtain flexural stiffness. In the first mode, the device reproduces the natural stalk lodging failure mode (buckling). Whereas the second mode of testing allows for a non-destructive measurement that is a surrogate for laboratory-based stalk bending strength measurements (Robertson et al., 2016; Cook et al., 2019). However, in the nondestructive mode the reliability of measurements depends on soil conditions and soil-type, which should ideally be kept constant throughout testing. A recent study which utilized the DARLING device as part of multi-year, multi-location study demonstrated that bending strength measurements are more highly correlated with natural lodging incidence as compared to rind penetration testing (Sekhon et al., 2020). However, the study also indicated that rind penetration resistance does account for part of the observed variability in natural lodging incidence that is not accounted for by bending strength measurements.

# METHODS TO EVALUATE ROOT LODGING

### *ROOT PULLING/PUSHING RESISTANCE*

While the failure mechanics of roots during lodging includes both uprooting and breakage, field-based approaches have focused on measuring plant anchorage independent of these failure mechanics. Vertical root pulling resistance (VRPR) is a parameter for assessing root anchorage that has been widely used in maize since the 1930s (Wilson, 1930; Zuber et al., 1971; Fincher et al., 1985). VRPR can be measured rapidly in the field and was shown to be negatively correlated with root lodging in maize (Kamara et al., 2003; Liu et al., 2011). VRPR has been less utilized in other cereal crops, particularly in the context of root lodging. One set of studies used VRPR in rice as an approach to understand and select for drought-tolerance, but did not evaluate root lodging (Ekanayake et al., 1985; Ekanayake et al., 1986). Another study determined that VRPR was highly variable and unsatisfactory to predict the tendency to root lodge in wheat, oats, and barley (Harrington and Waywell, 1950). Despite the unclear relationship with root lodging resistance, several devices have been developed to measure VRPR in the field.

An early device to measure VRPR consisted of a clamp and a scale, where the plant is lifted from the soil by manually pushing a lever; this method has proven inaccurate since it was impossible to control for lifting rate and measurements were extracted manually (Rogers et al., 1976; Thompson, 1972, 1982; Jenison et al., 1981; Penny, 1981; Arihara and Crosbie, 1982; Peters et al., 1982). Other devices have been subsequently designed to reduce manual error and measure VRPR using tractor hydraulics, but this approach has proven too heavy and cumbersome for wide-spread measurements (Zuber, 1968; Donovan et al., 1982; Kevern and Hallauer, 1983; Melchinger et al., 1986).

Two devices were designed to overcome the limitations identified from these early methods to measure VRPR (Dourleijn et al., 1988; Fouéré et al., 1995). One device (referred to here as Dourleijn's Device; Fig.4A, Table 2) uses an electric powered motor and pulley system to pull the plants vertically out of the soil at a constant rate (Dourleijn et al., 1988). The maximum pulling force is recorded as a post-test on an attached scale. A second device (referred to here as Fouéré's Device; Fig.4B, Table 2) is anchored by nails into the soil and the stalk is symmetrically placed between the anchor feet (Fouéré et al., 1995). A force sensor then transmits an angular pushing displacement to the stalk and records the resistance force at discrete angles. Mechanical data are recorded as moment-angle relationships and the maximum force applied to pull the root system out of the soil is then extracted as the horizontal root pushing resistance. This device represented several improvements upon previous devices including the use of fork prongs to prevent root system damage, nails to anchor the device to the soil, and automated recording of force measurements.

## *ROOT FAILURE MOMENT*

A major drawback of the root pulling/pushing systems is that they do not necessarily replicate how a plant fails during root lodging. In other words, they apply unnatural loads and likewise produce unnatural failure types and patterns. An attempt to improve upon these approaches and replicate root lodging conditions was made with the introduction of a device to measure root failure moment (Rfm). This approach was originally designed for sunflower (Sposaro et al., 2008), and subsequently applied to maize (Liu et al., 2012). The devices to measure Rfm consist of a push bar attached to the plant stem at a specific height with a steel cable, a base protractor and an offset pulley system to pull the plant over (referred to here as Sposaro's Device; Fig. 4C, Table 2). The Rfm is then calculated as the force when the stalk is pulled to perpendicular multiplied by the attachment height of the push bar. While not widely used (likely due to the cumbersome and low-throughput nature of the device), Rfm in maize was shown to be negatively correlated with planting density, which is known to increase root lodging (Liu et al., 2012).

### HANDHELD PROSTRATE TESTING

An approach to measure root anchorage in small grains has been the use of a commercially available, handheld prostrate testing device (Daiki Rika Kogyo Co., Ltd, Saitama, Japan). In this system, the prostrate device is attached perpendicular to multiple plant stems (10-15), the plants are displaced to a 45-degree angle and the pushing resistance is recorded (Fig. 4D, Table 2). This approach has been applied to winter wheat (Xiao et al., 2015), canola (Wu and Ma, 2016), and rice (Kashiwagi and Ishimaru, 2004). Interestingly, this approach is very similar to the bending tests used to assess stalk lodging, varying only in the placement of the device lower on the stem and closer to the soil surface. While this approach is often presented as a measure of root anchorage, one study notes that it is difficult to separate this measure as indicative of root lodging distinct from stalk lodging (Xiao et al., 2015). For example, in rice the bending moment calculated from prostrate testing of the lowest internodes was highly predictive of Blaster-induced stalk lodging (Shrestha et al., 2020).

## ALTERNATIVE ANALYSES OF BIOMECHANICS

There are two primary alternatives to field-based measurements: laboratory-based measurements and computational modeling. While these topics are outside the primary focus of this review, the basic features of these alternatives are presented as a starting point for further reading.

Laboratory-based measurements rely on samples being removed from the field and transported to a laboratory. These types of analyses for stalks include destructive crushing tests, bending tests, or analyses of plant anatomy. A recent review provides an overview of the laboratory-based measurements of stalk mechanics (Shah et al., 2017), thus they are not discussed here. In contrast to stalk lodging, there are limited approaches that have been used to understand root lodging in the laboratory setting due to the root system being less amenable to removal and mechanical testing.

One attempt to understand root lodging outside of the field setting was the use of computational models of root/soil interactions that were used to gain new insights on the factors influencing root lodging in maize (Brune et al., 2018). These models allow researchers to explore hypotheses and carry out "computational experiments" that could not be accomplished with purely empirical approaches. One major advantage of computational models is that every aspect of a computational model can be independently manipulated. This enables experiments that are fundamentally different from the experiments that can be performed in either the laboratory or the field. For example, computational modeling has been used to dissect the factors influencing stalking lodging in maize (von Forrell et al., 2015). Additional information on computational modeling to understand plant biomechanics can be found in a review article by Pursinkiewicz and Runions (2012).

## DISCUSSION

A general overview of field-based mechanical phenotyping equipment used to assess lodging resistance in grain crops is provided in Tables 1-2. In the sections below, we discuss the need for improved standard operating procedures / testing standards for phenotyping equipment that will enable greater interoperability. In addition, we provide an evaluation of each phenotyping method and mention best practices for conducting field based mechanical tests of plant stalks and roots.

### Development of Testing Standards to Enable Reproducibility

As technologies to assess plant mechanics in a field setting continue to be developed, there is an urgent need to focus on reproducibility and a complete understanding of the mechanics of plant failure. Several of the approaches outlined above suffer from a lack of reproducibility between laboratories and/or devices. This lack of reproducibility comes in part from a failure to understand how plants fail during lodging. For example, there is little conceptual relationship between rind puncture resistance (pushing a needle-like instrument laterally through the outer tissues of the stalk) and natural wind-induced failure, which typically manifests as buckling, snapping, or splitting (Robertson et al., 2015a). Similarly, for root lodging, plants are not pulled vertically from the soil during natural lodging events. It thus follows that root pulling/pushing tests have shown variable success to understand root lodging resistance.

Another challenge in reproducibility is the lack of experimental detail included in publications. It is often unclear how devices are constructed, how they are used and/or what mechanical metrics are measured. A prime example of this is the copious use of the term "strength", which is meaningless without context. Two contexts are of importance when discussing strength, the first is concerning how strength was measured (e.g. bending, crushing, shear, tensile, compressive, etc.) and the second with how strength was extracted. For example, there are often two strength measures associated with plant mechanics - the ultimate structural strength and the structural failure or yield strength. The ultimate structural strength represents the highest force that the object can withstand, whereas the structural failure or yield strength represents the force at which the object breaks or buckles. These measures are not always equivalent, and should be differentiated when reporting results. Furthermore, structural strength should be differentiated from material strength. The ultimate structural strength of an object is the highest force it can withstand and does not account for differences in geometry among objects. Whereas, the ultimate material strength is the highest force per unit area a given material can withstand. The ultimate structural strength is the type of strength most commonly measured in plant phenotyping experiments. Clarifying the type of strength measured by devices in future manuscripts will enable greater interoperability and understanding.

Lastly, there remains a lack of connection between the field-based mechanical measures described here and underlying biology. Linking field-based mechanical measures to plant anatomy, architecture, and composition is a key next phase of research in mechanical phenotyping. Understanding how mechanical measures vary with the underlying biology enhance our ability to select for plants with improved lodging resistance without compromising other traits such as yield or disease-resistance.

### Evaluation of Phenotyping Methods

In this section, we endeavor to provide an objective evaluation of current phenotyping methods. This evaluation is based on the authors' experience, opinion and fairly limited data availability. It is intended to assist plant scientist in choosing a phenotyping method and to highlight potential future research directions. A ranking of phenotyping devices is not provided as our intent is only to provide an objective evaluation of each methodology. Each method has its own unique benefits and drawbacks as discussed below.

As compared to other phenotyping methods, artificial wind is generally assumed to most closely mimic naturally applied forces that induce stalk and root lodging. Although artificial wind has been shown to induce stalk lodging in rice with a high correlation to natural lodging (Shrestha, et al., 2020), we are unable to find any literature stating that these machines do or do not produce natural stalk lodging or root lodging failure types and patterns. Additionally, the large cost and size of such machines make them inaccessible to many public-sector plant breeders / agronomists. The limited mobility (with the exception of Boreas) also makes wind machines difficult to utilize on large association panels to discover the genetic underpinning of lodging resistance. A primary limitation of artificial wind machines that are static relative to the field is that wind speed naturally decreases with distance from the wind source. Thus, plants near the wind source will experience different wind forces than those near the back of a plot. Various correction factors or other manners of accounting for this have been presented in the referenced articles. However, the Boreas machine has overcome this limitation altogether and is likely the most well developed artificial wind source. This is due to a large corporate investment in the machine, however, as mentioned previously, the Boreas is inaccessible to most breeders and most of the data collected with the machine is proprietary. To increase accessibility, future studies should strive to lower the cost and improve mobility of artificial wind sources. In addition, future researchers should seek to confirm that artificial wind induces natural failure types and patterns in cereal crops to confirm the validity of use as a proxy for lodging resistance.

Rind penetration experiments have been conducted for nearly 100 years, yet much is still unknown about the methodology. For example, it is unclear how probe geometry and rate of force application affect the measurement. A primary advantage of rind penetration is that it is not entirely destructive (i.e. it does induce plant death) and the testing can be done before flowering. This enables plant scientist to make breeding decisions (e.g. experiment crosses etc.) in the same season that the data is collected. This is not possible when utilizing artificial wind sources or other destructive phenotyping measurements of lodging resistance. Rind penetration testing is also one of the most rapid ways of mechanically phenotyping cereal crops, and enables testing any given plant without disturbing neighboring plants. This method is effective for rapidly ranking varieties with significant differences in stalk strength, but does not perform well at differentiating between elite varieties, which may possess very similar strengths. The primary weakness of this approach is that rind penetration experiments do not produce natural stalk lodging failure patterns. Thus, breeding for increased rind penetration resistance may not always increase stalk lodging resistance. The lack of association with natural failure types and patterns is one likely reason that previous studies have demonstrated mixed results. We are unaware of any studies using rind penetration resistance to study small grains. The puncture force of small grains is likely so low as to complicate accurate measurement in a field setting. In summary rind penetration resistance appears to be a viable way to rapidly investigate stalk lodging resistance, but additional factors (e.g. plant geometry or bending strength) also need to be considered when breeding for stalk strength. Future studies should report the rate of force application and give a detailed description of the probe geometry used in the study. In addition, there is a need for the field to settle on a standard operating procedure for rind penetration resistance testing.

Bending tests can induce the same failure patterns observed in naturally lodged crops. At the time of writing, the DARLING device is the only field-based bending strength device explicitly shown to produce natural failure types and patterns (Cook et al, 2019). However, we believe it is highly probable that the other bending strength devices (which operate on similar principles) likewise produce natural failure types and patterns. The bending test approach essentially eliminates the chaotic influence of wind loading, thus decreasing measurement uncertainty. In other words, these tests provide information about the inherent strength of individual stalks in the absence of wind effects. For example, artificial wind tests are affected by leaf size, leaf number, leaf angle, planting density etc. From a statistical standpoint the bending strength approach likely provides the greatest distinguishing power for ranking varieties based on inherent stalk strength. Of course, these methods are not without their own unique drawbacks. The throughput of bending tests can vary, but in general this method typically takes longer than the rind puncture test (testing rates of most wind tests are not well-quantified). The DARLING device appears to be the fastest of the bending test methods, with testing rates of approximately 150 – 200 stalks/hour. The results of bending tests must be interpreted with caution. While the method does provide very detailed information about stalk strength, it does not provide any information about how each variety translates wind into bending load. Thus, one can imagine a situation in which a crop variety with a high strength rating may lodge at a higher rate than a similar variety with a slightly lower strength, because the leaf architecture of the first variety results in higher bending loads than the leaf architecture of the second variety.

All devices utilized for testing small grains suffer from one common limitation or constraint - they require testing of multiple plants at once. During the test these plants interact with one another and can provide mechanical support to one another. The physics governing this self-supporting behavior are complex and have not been fully elucidated. To overcome this limitation, the measured force is typically divided by the number of stems deflected in the test to develop an average strength value. It is unclear how different plant spacings, or the number of tillers etc. may affect the average strength value.

In general there has been greater progress in understanding the mechanics associated with stalk lodging than the mechanics associated with root lodging. Root lodging is more difficult to simulate due to a lack of devices that apply a rotational moment at the base of the plant. Specifically, root pushing/pulling devices have suffered from a lack of reproducibility and variable predictability of natural root lodging. The failure pattern of roots achieved using these devices has not been reported, thus it remains unclear if variation in predictability is related to the type of failure induced or the wrong direction of applied force to mimic root lodging.

Devices that measure root failure moment are likely the most appropriate to determine root lodging susceptibility, but are currently cumbersome and low-throughput. Further, the one study that reported results in maize did not detail the failure pattern of the roots nor provide a direct comparison with natural root lodging (Liu et al., 2012). There is an urgent need to develop phenotyping approaches that simulate the natural failure patterns during root lodging. However, the natural failure patterns during root lodging are poorly understood. It is likely that development of devices that apply a rotational moment at the base of the plant will enable a better understanding of the root-soil interactions that are critical for plant anchorage. It should also be noted that several patents exist for devices to measure crop lodging resistance.

However, the majority of these devices are not reported in scientific literature and their efficacy is therefore not possible to assess in this review.

Overall, the devices discussed in this review could be made more user-friendly by supporting the collection of metadata (plot #, gps coordinates etc.) and utilizing improved digital user interfaces. These features are necessary to make the devices amenable to large genetic studies, to limit inter-user variability, data corruption, and data loss. Further, the accessibility of field-based mechanical phenotyping equipment is limited. The technology is rapidly advancing, and the field lacks a consensus on the best phenotyping approaches. In our experience, the best practice for acquiring field-based mechanical phenotyping data is to collaborate directly with the developers of the instrument(s) being used. Data collection pitfalls are many and are unique to each device, crop, and field combination. Integrative research teams combining expertise in plant science and mechanical measurements are required to successfully identify and navigate such pitfalls in a field-situation.

## ACKNOWLEDGMENTS

This work was supported by grants from the National Science Foundation (#1826715), and the USDA-National Institute of Food and Agriculture (#2016-67012-28381) to DJR. This work was supported by grants from the Delaware Biosciences Center for Advanced Technology, the University of Delaware Research Foundation, and the Thomas Jefferson Fund / FACE Foundation to EES. Any opinions, findings, conclusions, or recommendations are those of the author(s) and do not necessarily reflect the view of the funding bodies.

## AUTHOR CONTRIBUTIONS

All authors contributed to the writing of the manuscript and approved the final version. LE illustrated all figures.

**LITERATURE CITED**

Anderson, B., and D. G. White. 1994. "2430801. Evaluation of Methods for Identification of Corn Genotypes with Stalk Rot and Lodging Resistance." *Plant Disease* 78 (6): 590–93.

Arihara, J., and T. M. Crosbie. 1982. "Relationships among Seedling and Mature Root System Traits of Maize 1." *Crop Science* 22 (6): 1197–1202.

Baker, C. J. 1995 . "The Development of a Theoretical Model for the Windthrow of Plants." *Journal of Theoretical Biology* 175: 355-72.

Barrerio, R., L. Carrigan, M. Ghaffarzadeh, D. M. Goldman, M. E. Hartman, D. L. Johnson, and L. Steenhoek. 2008. Device and method for screening a plant population for wind damage resistance traits. USPTO 7412880B2. *US Patent*, filed October 13, 2006, and issued August 9, 2008.

Berry, P. M., J. H. Spink, A. P. Gay, and J. Craigon. 2003a. "A comparison of root and stem lodging risks among winter wheat cultivars." *Journal of Agricultural Science* 141: 191-202.

Berry, P. M., J. Spink, and M. Sterling. 2003b. "Methods for Rapidly Measuring the Lodging Resistance of Wheat Cultivars." *Journal of Agronomy & Crop Science* 189: 390-401.

Berry, P. M., M. Sterling, J. H. Spink, C. J. Baker, R. Sylvester-Bradley, S. J. Mooney, A. R. Tams, and A. R. Ennos. 2004. "Understanding and Reducing Lodging in Cereals." In *Advances in Agronomy*, 84:217–71.

Berry, P. M. 2013. "Lodging Resistance in Cereals." In P. Chirstou, R. Savin, B. A. Costa-Pierce, I. Misztal, and C. B. A. Whitelaw (eds) *Sustainable Food Production*. Springer, New York, NY.

Brazier, L. G. 1927. "On the Flexure of Thin Cylindrical Shells and other "Thin" sections." *Proceedings of the Royal Society A* 116 (773): 104-14.

Brune, P. F., A. Baumgarten, S. J. McKay, F. Technow, and J. J. Podhiny. 2018. "A Biomechanical Model for Maize Root Lodging." *Plant and Soil* 422 (1): 397–408.

Butrón, A., R. A. Malvar, P. Revilla, P. Soengas, A. Ordás, and H. H. Geiger. 2002. "Rind Puncture Resistance in Maize: Inheritance and Relationship with Resistance to Pink Stem Borer Attack." *Plant Breeding* 121 (5): 378–82.

Carter, P. R., and K. D. Hudelson. 1988. "Influence of Simulated Wind Lodging on Corn Growth and Grain Yield." *Journal of Production Agriculture* 1: 295–99.

Cook, D. D., W. de la Chapelle, T.-C. Lin, S. Y. Lee, W. Sun, and D. J. Robertson. 2019. "DARLING: a device for assessing resistance to lodging in grain crops" *Plant Methods* 15: 102.

Cooper, M., C. D. Messina, D. Podlich, L. Radu Totir, A. Baumgarten, N. J. Hausmann, D. Wright, and G. Graham. 2014. "Predicting the future of plant breeding: complementing empirical evaluation with genetic prediction." *Crop & Pasture Science* 65: 311-36.

Crook, M. J., and A. R. Ennos. 1993. "The Mechanics of Root Lodging in Winter Wheat, Triticum Aestivum L." *Journal of Experimental Botany* 44 (264): 1219–24.

Crook, M. J., and A. R. Ennos. 1994. "Stem and Root Characteristics Associated with Lodging Resistance in Four Winter Wheat Cultivars." *The Journal of Agricultural Science* 123 (2): 167–74.

Donovan, L. S., P. Jui, M. Kloek, and C. F. Nicholls. 1982. "An Improved Method Of Measuring Root Strength In Corn (Zea Mays L.)." *Canadian Journal of Plant Science*.

Dourleijn, C. J., A. P. M. den Nijs, and O. Dolstra. 1988. "Description and Evaluation of a Device for Measuring Vertical Pulling Resistance in Maize (Zea Mays L.)." *Euphytica*.

Dudley, J. W. 1994. "Selection for Rind Puncture Resistance in Two Maize Populations." *Crop Science* 34: 1458–60.

Easson, D. L., E. M. White, and S. J. Pickles. 1992. "A Study of Lodging in Cereals." *HGCA Research Project No. 52. Home-Grown Cereals Authority*.

Ekanayake, I. J., D. P. Garrity, T. M. Masajo, and J. C. O'Toole. 1985. "Root Pulling

Resistance in Rice: Inheritance and Association with Drought Tolerance." *Euphytica*.

Ekanayake, I. J., D. P. Garrity, and J. C. O'Toole. 1986. "Influence of Deep Root Density on Root Pulling Resistance in Rice1." *Crop Science*.

Elmore, R. W., D. B. Marx, R. G. Klein, and L. J. Abendroth. 2005. "Wind Effect on Corn Leaf Azimuth." *Crop Science* 45 (6): 2598–2604.

Ennos, A. R., M. J. Crook, and C. Grimshaw. 1993. "The Anchorage Mechanics of Maize, Zea mays." *Journal of Experimental Botany* 44 (258): 147-153.

Fincher, R. R., L. L. Darrah, and Zuber, M. S. 1985. "Root Development in Maize as Measured by Vertical Pulling Resistance." *Maydica* 30.

Flint-Garcia, S. A., C. Jampatong, L. L. Darrah, and M. D. McMullen. 2003a. "Quantitative Trait Locus Analysis of Stalk Strength in Four Maize Populations." *Crop Science*.

Flint-Garcia, S. A., M. D. McMullen, and L. L. Darrah. 2003b. "Genetic Relationship of Stalk Strength and Ear Height in Maize." *Crop Science*.

von Forell, G., D. J. Robertson, S. Y. Lee, and D. D. Cook. 2015. "Preventing Lodging in Bioenergy Crops: A Biomechanical Analysis of Maize Stalks Suggests a New Approach." *Journal of Experimental Botany* 66 (14): 4367–71.

Fouéré, A., S. Pellerin, and A. Duparque. 1995. "A Portable Electronic Device for Evaluating Root Lodging Resistance in Maize." *Agronomy Journal* 87: 1020–24.

Gardiner, B., P. Berry, and B. Moulia. 2016. "Review: Wind Impacts on Plant Growth, Mechanics and Damage." *Plant Science: An International Journal of Experimental Plant Biology* 245 (April): 94–118.

Gibson, B. K., C. D. Parker, and F. R. Musser. 2010. "Corn Stalk Penetration Resistance as a Predictor of Southwestern Corn Borer (Lepidoptera: Crambidae) Survival." *Midsouth Entomologist* 3: 7–17.

Grafius J. E., and H. M. Brown. 1954. "Lodging Resistance in Oats." *Agronomy Journal* 46 (9): 414-8.

Gou, L., J. Huang, B. Zhang, T. Li, R. Sun, and M. Zhao. 2007. "Effects of Population Density on Stalk Lodging Resistant Mechanism and Agronomic Characteristics of Maize."

Guo, Q., R. Chen, X. Sun, M. Jiang, H. Sun, S. Wang, L. Ma, Y. Yang, and J. Hu. 2018. "A Non-Destructive and Direction-Insensitive Method Using a Strain Sensor and Two Single Axis Angle Sensors for Evaluating Corn Stalk Lodging Resistance." *Sensors* 18 (6).

Guo, Q., R. Chen, L. Ma, H. Sun, M. Weng, S. Li, and J. Hu. 2019. "Classification of Corn Stalk Lodging Resistance Using Equivalent Forces Combined with SVD Algorithm." *Applied Sciences*, *9* (4): 640.

Harrington, J. B., and C. G. Waywell. 1950. "Testing Resistance to Shattering and Lodging in Cereals." *Scientific Agriculture* 30 (2): 51–60.

Hu, H., Y. Meng, H. Wang, H. Liu, and S. Chen. 2012. "Identifying Quantitative Trait Loci and Determining Closely Related Stalk Traits for Rind Penetrometer Resistance in a High-Oil Maize Population." *Theoretical and Applied Genetics.* 124 (8): 1439–47.

Jenison, J. R., D. B. Shank, and L. H. Penny. 1981. "Root Characteristics of 44 Maize Inbreds Evaluated in Four Environments 1." *Crop Science* 21 (2): 233–37.

Jo Heuschele, D., J. Wiersma, L. Reynolds, A. Mangin, Y. Lawley, and P. Marchetto. 2019. "The Stalker: An Open Source Force Meter for Rapid Stalk Strength Phenotyping." *HardwareX* 6 (October): e00067.

Kamara, A. Y., J. G. Kling, A. Menkir, and O. Ibikunle. 2003. "Association of Vertical Root-Pulling Resistance with Root Lodging and Grain Yield in Selected S1 Maize Lines Derived from a Tropical Low-Nitrogen Population." *Journal of Agronomy and Crop Science* 189 (3): 129–35.

Kashiwagi, T., and K. Ishimaru. 2004. "Identification and Functional Analysis of a Locus for Improvement of Lodging Resistance in Rice." *Plant Physiology* 134 (2): 676–83.

Kevern, T. C., and A. R. Hallauer. 1983. "Relation of Vertical Root-Pull Resistance and Flowering in Maize." *Crop Science* 23 (2): 357–63.

Khanna, K. L. 1935. "An Improved Instrument for Testing Rind Hardness in Sugarcane." *Agr. and Live Stock in India* 5: 156–58.

Khobra, R., S. Sareen, B. K. Meena, A. Kumar, V. Tiwari, and G. Singh. 2018. "Exploring the traits for lodging tolerance in wheat genotypes: a review." *Physiology and Molecular Biology of Plants* 25: 589–600

Laude, H. H., and A. W. Pauli. 1956. "Influence of Lodging on Yield and Other Characters in Winter Wheat." *Agronomy Journal*.

Li, K., J. Yan, J. Li, and X. Yang. 2014. "Genetic Architecture of Rind Penetrometer Resistance in Two Maize Recombinant Inbred Line Populations." *BMC Plant Biology* 14 (June): 152.

Liu, J., H. Cai, Q. Chu, X. Chen, F. Chen, L. Yuan, G. Mi, and F. Zhang. 2011. "Genetic Analysis of Vertical Root Pulling Resistance (VRPR) in Maize Using Two Genetic Populations." *Molecular Breeding: New Strategies in Plant Improvement* 28 (4): 463–74.

Liu, S., F. Song, F. Liu, X. Zhu, and H. Xu. 2012. "Effect of Planting Density on Root Lodging Resistance and Its Relationship to Nodal Root Growth Characteristics in Maize (Zea Mays L.)." *Journal of Agricultural Science* 4 (12): 182.

Ma, D., R. Xie, X. Liu, X. Niu, P. Hou, K. Wang, Y. Lu, and S. Li. 2014. "Lodging-Related Stalk Characteristics of Maize Varieties in China since the 1950s." *Crop Science* 54: 2805–14.

Martin, S. A., L. L. Darrah, and B. E. Hibbard. 2004. "Divergent Selection for Rind Penetrometer Resistance and Its Effects on European Corn Borer Damage and Stalk Traits in Corn." *Crop Science* 44: 711–17.

McRostie, G. P., and J. D. MacLachlan. 1942. "Hybrid Corn Studies I." *Scientific Agriculture* 22 (5): 307–13.

Melchinger, A. E., G. A. Schmidt, and H. H. Geiger. 1986. "Evaluation of Near Infra-Red Reflectance Spectroscopy for Predicting Grain and Stover Quality Traits in Maize." *Plant*

*Breeding*.

Mizuno, H., S. Kasuga, and H. Kawahigashi. 2018. "Root Lodging Is a Physical Stress That Changes Gene Expression from Sucrose Accumulation to Degradation in Sorghum." *BMC Plant Biology* 18 (1): 2.

Moulia, B. 2013. "Plant biomechanics and mechanobiology are convergent paths to flourishing interdisciplinary research." *Journal of Experimental Botany* 64 (16): 4617-33.

Mulder, E. G. 1954. "Effect of Mineral Nutrition on Lodging of Cereals." *Plant and Soil* 5 (3): 246–306.

Neenan, M., and J. L. Spencer-Smith. 1975. "An Analysis of the Problem of Lodging with Particular Reference to Wheat and Barley." *The Journal of Agricultural Science* 85 (3): 495–507.

Niklas, K.J. 1992. *Plant biomechanics: an engineering approach to plant form and function*. University of Chicago press.

Niklas, K. J., and H. C. Spatz. 2012. *Plant physics*. University of Chicago Press.

Niu, L., S. Feng, W. Ding, and G. Li. 2016. "Influence of speed and rainfall on large-scale wheat lodging from 2007 to 2014 in China." *PLoS ONE* 11 (7): e0157677.

Ookawa, T., and K. Ishihara. 1992. "Varietal difference of physical characteristics of the culm related to lodging resistance in paddy rice." *Japanese Journal of Crop Science* 61: 419-25.

Peiffer, J. A., S. A. Flint-Garcia, N. De Leon, M. D. McMullen, S. M. Kaeppler, and E. S. Buckler. 2013. "The Genetic Architecture of Maize Stalk Strength." *PLoS One* 8 (6): e67066.

Penny, L. H. 1981. "Vertical-Pull Resistance of Maize Inbreds and Their Testcrosses." *Crop Science* 21 (2): 237–40.

Peters, D. W., D. B. Shank, and W. E. Nyquist. 1982. "Root-Pulling Resistance and Its Relationship to Grain Yield in F1 Hybrids of Maize1." *Crop Science*.

Prusinkiewicz, P., and A. Runions. 2012. "Computational models of plant development and form." *New Phytologist* 193(3): 549-69.

Rajkumara, S. 2008. "Lodging in Cereals-A Review." *Agricultural Reviews* 29 (1): 55.

Robertson, D. J., M. Julias, B. W. Gardunia, T. Barten, and D. D. Cook. 2015a. "Corn Stalk Lodging: A Forensic Engineering Approach Provides Insights into Failure Patterns and Mechanisms." *Crop Science* 55: 2833–41.

Robertson, D. J., S. L. Smith, and D. D. Cook. 2015b. "On Measuring the Bending Strength of Septate Grass Stems." *American Journal of Botany* 102 (1): 5–11.

Robertson, D. J., S, Smith, B. Gardunia, and D. D. Cook. 2014. "An Improved Method for Accurate Phenotyping of Corn Stalk Strength." *Crop Science* 54: 2038–44.

Robertson, D. J., S. Y. Lee, M. Julias, and D. D. Cook. 2016. "Maize Stalk Lodging: Flexural Stiffness Predicts Strength." *Crop Science*.

Robertson, D. J., M. Julias, S. Y. Lee, and D. D. Cook. 2017. "Maize Stalk Lodging: Morphological Determinants of Stalk Strength." *Crop Science* 57: 926–34.

Rogers, R. R., W. A. Russell, and J. C. Owens. 1976. "Evaluation of a Vertical-Pull Technique in Population Improvement of Maize for Corn Rootworm Tolerance 1." *Crop Science* 16 (4): 591–94.

Sekhon, R. S., C. N. Joyner, A. J. Ackerman, C. S. McMahan, D. D. Cook, and D. J. Robertson. 2020. "Stalk Bending Strength is Strongly Associated with Maize Stalk Lodging Incidence Across Multiple Environments." *Field Crops Research* 249 (1): 107737.

Shah, D. U., T. P. S. Reynolds, M. H. Ramage, and C. Raines. 2017. "The strength of plants: theory and experimental methods to measure the mechanical properties of stems." *Journal of Experimental Botany* 68 (16): 4497-516.

Shrestha, S., M.R.C. Laza, K.V. Mendez, S. Bhosale, and M. Dingkuhn. 2020 "The blaster: A methodology to induce rice lodging at plot scale to study lodging resistance." *Field Crops Research* 245.

Sposaro, M. M., C. A. Chimenti, and A. J. Hall. 2008. "Root Lodging in Sunflower. Variations in Anchorage Strength across Genotypes, Soil Types, Crop Population Densities and Crop Developmental Stages." *Field Crops Research* 106 (2): 179–86.

Stamp, P., and C. Kiel. 1992. "Root Morphology of Maize and Its Relationship to Root Lodging." *Journal of Agronomy and Crop Science* 168 (2): 113–18.

Sterling, M., C. J. Baker, P. M. Berry, and A. Wade. 2003. "An experimental investigation of the lodging of wheat." *Agricultural and Forest Meteorology* 119: 149-65.

Stubbs, C. J., Cook D. D., and K. J. Niklas. 2019. "A general review of the biomechanics of root anchorage." *Journal of Experimental Botany* 70 (14): 3439-51.

Thompson, D. L. 1972. "Recurrent Selection for Lodging Susceptibility and Resistance in Corn 1." *Crop Science* 12 (5): 631–34.

Thompson, D. L. 1982. "Grain Yield of Two Synthetics of Corn after Seven Cycles of Selection for Lodging Resistance 1." *Crop Science* 22 (6): 1207–10.

Wen, W., S. Gu, B. Xiao, C. Wang, J. Wang, L. Ma, Y. Wang, et al. 2019. "In Situ Evaluation of Stalk Lodging Resistance for Different Maize (Zea Mays L.) Cultivars Using a Mobile Wind Machine." *Plant Methods* 15 (August): 96.

White, E. M. 1991. "Response of Winter Barley Cultivars to Nitrogen and a Plant Growth Regulator in Relation to Lodging." *The Journal of Agricultural Science*.

Wilson, H. K. 1930. "Plant Characters as Indicies in Relation to the Ability of Corn Strains to Withstand Lodging." *Journal of the American Society of Agronomy* 22.

Wu, W., and B.-L. Ma. 2016. "A New Method for Assessing Plant Lodging and the Impact of Management Options on Lodging in Canola Crop Production." *Scientific Reports* 6 (August): 31890.

Xiao, Y., J. Liu, H. Li, X. Cao, X. Xia, and Z. He. 2015. "Lodging Resistance and Yield Potential of Winter Wheat: Effect of Planting Density and Genotype." *Frontiers of Agricultural Science and Engineering* 2 (2): 168.

Zuber, M. S. 1968. "Evaluations of Corn Root Systems under Various Environments." In *Amer Seed Trade Assoc Hybrid Corn Indus Res Conf Proc.*

Zuber, M. S., G. J. Musick, and M. L. Fairchild. 1971. "A Method of Evaluating Corn Strains for Tolerance to the Western Corn Rootworm." *Journal of Economic Entomology* 64 (6): 1514–18.

## TABLES

| Table 1. Comparison of Stalk Mechanical Phenotyping Devices | | | | |
|---|---|---|---|---|
| **Device** | **Crop** | **Measurement** | **Pros** | **Cons** |
| **Berry's Device** | Wheat (multiple plants at once) | Stem failure strength<br><br>Root failure strength (with soil wetting) | • Inexpensive<br>• Adjustable push-bar height<br>• Digital force-torque display<br>• Rapid measurements (6 min/plot) | • Unable to measure individual plant properties<br>• Heavy battery pack<br>• Results dependent on soil conditions |
| **Stalker** | Wheat, oat, barley | Resistance force | • Open Source<br>• Nondestructive<br>• Field-portable<br>• Can test singular or multiple plants<br>• LED indicates maximum rotation at 45 degrees.<br>• Data stored on SD card<br>• Continuous force-displacement data collection | • No digital display<br>• Before data analysis, load values must be converted to units of Newtons using calibration curve<br>• Only used for small grains<br>• Operator dependent data (prone to jitter) |
| **Guo's Device** | Maize | Pulling force<br><br>Equivalent force | • Nondestructive<br>• Portable<br>• Digital display<br>• Data stored in flash memory<br>• Flexible pulling directions | • Difficult to translate measurement to engineering principles<br>• Operator dependent data (prone to jitter) |
| **DARLING** | Maize, sorghum | Failure strength<br><br>Flexural stiffness | • Replicates natural failure patterns<br>• Field-portable<br>• Rapid measurements (17 sec/test)<br>• Can be nondestructive<br>• Adjustable load cell<br>• Graphical user interface<br>• Continuous force-displacement data collection | • Only useful for large grains<br>• Accuracy of measurements decreases as rotation increases past 20 degrees<br>• Dependent on soil conditions<br>• Operator dependent data (prone to jitter) |

| Table 2. Comparison of Root Mechanical Phenotyping Devices | | | | |
|---|---|---|---|---|
| **Device** | **Crop** | **Measurement** | **Pros** | **Cons** |
| **Dourleijn's Device** | Maize | Vertical root pulling resistance | • Constant pulling speed (automated)<br>• Rapid measurements (5-10 mins/10 plants) | • Scale must be read and recorded manually<br>• Heavy (must be rolled like a wheelbarrow)<br>• Measurements are device specific<br>• Requires two operators<br>• Does not mimic failure during root lodging |
| **<u>Fouéré's</u> Device** | Maize | Horizontal root pulling resistance | • Digital display<br>• Field-portable<br>• Data saved internally<br>• Automated recording of force measurements<br>• Rapid measurements (1 min/plant) | • Expensive ($5000 USD)<br>• Dependent on soil conditions<br>• Does not mimic failure during root lodging |
| **Sposaro's Device** | Maize, Sunflower | Root failure moment | • Inexpensive<br>• Adjustable push-bar height<br>• Field-portable<br>• Mimics failure during root lodging | • No internal memory<br>• Requires large row spacing to function. |
| **Prostrate Testing Device** | Small grains (canola, wheat, rice) | Root failure moment | • Inexpensive and commercially available<br>• Handheld<br>• Non-destructive<br>• Can be used for a variety of plants | • No digital display<br>• Difficult to conduct measurements in wind<br>• Loading rate is operator dependent<br>• Cannot distinguish between root and stalk properties. |

## GLOSSARY OF TERMS

**Bending** – Also known as flexure. Refers to the instance where an external force is applied perpendicular to the longitudinal axis of a slender object. Bending tests can be used in the field or laboratory with different equipment to obtain mechanical properties or phenotypes. Field-based configurations are discussed in this review, and laboratory-based configurations are explained in Shah et al. 2017.

**Brazier buckling** – Refers to the mechanism by which hollow tubes fail in bending. In this mechanism, transverse shear, causing an ovalization of the tube's cross section. This results in a characteristic creasing or buckling failure pattern (Brazier 1927).

**Lodging** – The displacement of plants from their vertical stance (Rajkumara, 2008). Lodging can occur due to stalk or root failure. **Stalk lodging** occurs when plants are bent or broken at the stalk below the inflorescence. **Root lodging** occurs when roots are broken or pulled from the soil.

**Rind Penetration** – Measured force required to pierce the rind of a plant stalk. This test is typically conducted with a probe attached to a force gauge. This measurement has been used to assess stalk lodging susceptibility, but the results are conflicting about the utility of this measurement and relationship to stalk lodging.

**Strength** – An object's ability to resist forces before deformation or failure. In engineering, the term can refer to the localization of forces using integrative modeling or the collective forces applied to a structural member or object. For the scope of this paper, strength refers to the latter, also known as structural strength (Moulia, 2013). There are multiple types of strength measurements. **Bending strength** (also known as flexural strength) is an object's ability to resist bending before accruing plastic or permanent deformations. **Failure strength** is the force at which an object breaks. **Ultimate tensile strength** is the maximum force that can be applied to an object before failure. **Yield strength** is the force at which elastic deformation ends and plastic or permanent deformation begins.

**Root pulling/pushing resistance** – The amount of force required to overcome soil adhesion and uproot a plant. Root pulling resistance is applied vertically and pushing resistance is applied horizontally.

## FIGURE LEGENDS

**Figure 1. Failure patterns of cereal stalks.** Natural failure patterns of cereal stalks vary by the crop-type and age of the plant. (A) In small grains, the most common failure mechanism is buckling at the lower internodes. (B) However, in barley and oat, buckling of middle internodes or as high as the peduncle have been reported. (C) In contrast for large grains, such as maize, mid-season failure occurs in a green (or brittle) snapping pattern, with failure at the stem nodes. (D) For late-season maize lodging, failure is defined by brazier buckling of stem internodes, close to the node.

**Figure 2. Failure patterns of cereal roots.** Experimental evidence suggests that cearl crop roots act as tethers in tension or compression. During root lodging, these tethers may fail through (A) roots pulling out of the soil and/or (B) roots breaking in either tension (on the side where force is applied) or compression (on the side away from where the force is applied).

**Figure 3. Devices for measuring stalk bending strength.** Berry's device (A) was developed to study winter wheat stalk lodging and consists of a hand-held force meter with a load cell attached to a push bar that measures the resistance force to push over in multiple plants. An updated version of this device called the Stalker (B), was developed for small grains (wheat, oat and barley) and reduced the weight of Berry's device and introduced some automation. Specifically, the Stalker is pushed forward until a preset 45-degrees, and then the force-rotation data is continuously recorded until the test is ended by the operator. Guo's device (C) features a hand-held two-component circuit block system and measures the forces required to bend maize stalks across a set of discrete angles. One component, a controller module, contains a strain sensor, single-axis angle sensor, microcontroller, power supply module, a signal acquisition circuit, and a radio frequency transceiver. The second component consists of another radio frequency transceiver and single axis sensor. The two components are connected by a rigid belt, and the controller is pulled to discrete angles to measure the maximum equivalent force ($F_{eq}$), which is used to assess stalk lodging resistance. The Device for Assessing Resistance to Lodging in Grains (DARLING) (D) was developed to assess stalk biomechanics in larger cereal crops and more closely recreate natural failure patterns during stalk lodging. DARLING consists of a vertical support with a control box and digital display mounted at the top, a horizontal footplate attached by a hinge at the base, and an adjustable height load cell attached. Plants can be non-destructively bent within the linear-elastic range of the material to obtain flexural stiffness or displaced until failure and the maximum applied bending strength recorded.

**Figure 4. Devices for measuring root anchorage.** Dourjelin's device (A) measures vertical root pulling resistance in maize. The device uses an electric powered motor and pulley system to pull the plants out of the soil at a constant rate. Fouéré's device (B) measures horizontal root pushing force in maize and consists of a main frame, handle, adjustable force sensor, angle sensor, a two-pronged steel fork with anchoring nails, and control head with electronic display and keys. This device uses a force sensor to transmit an angular pushing force to the stalk and an electronic control system automatically records the resistance force. Sposaro's device (C) was originally developed for sunflower and later applied to maize to improve upon root pulling/pushing resistance devices, and better replicates the failure mode of root lodging. With this device, a push bar is attached to the plant stem, while a base protractor and an offset pulley system are used to pull the plant over. Root failure moment (Rfm) can then be calculated. For smaller crops (canola, wheat, rice) a commercially available prostrate testing device (D) can been used. The device attaches to an adjustable mounted plate attached to the plant. Plants are displaced to a 45-degree angle and the pushing resistance is recorded.

**A - Small grain bucking at lower internode**

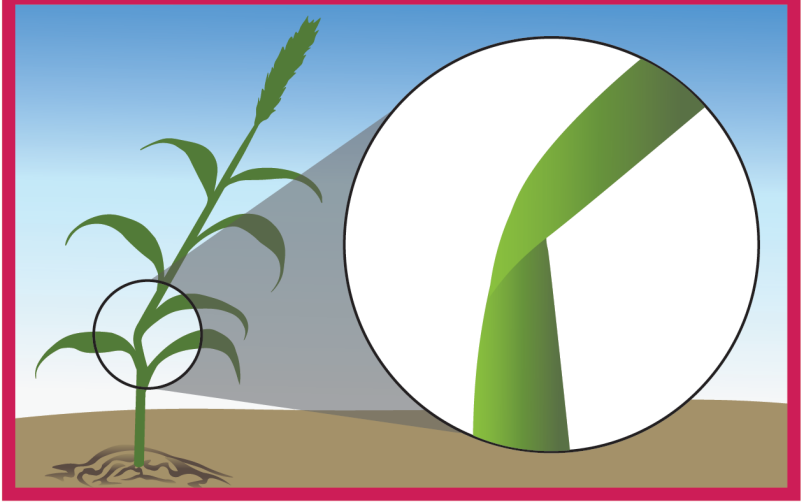

**B - Barley/oat buckling at middle internode and peduncle**

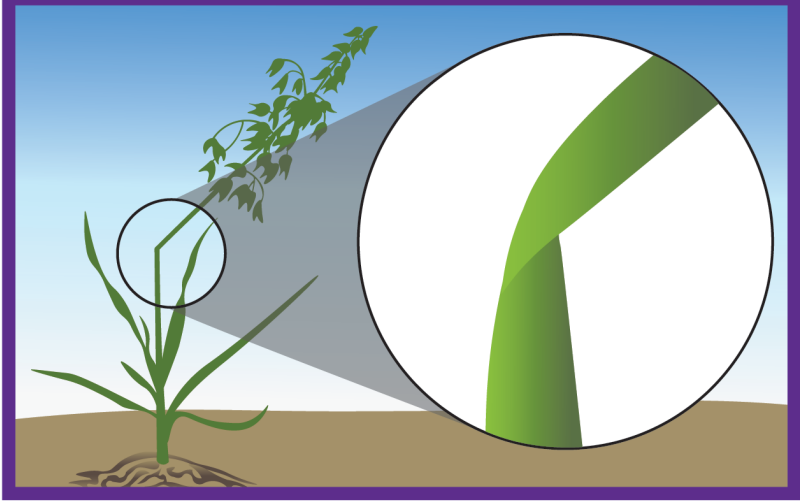

**C - Maize green snap at node (mid-season)**

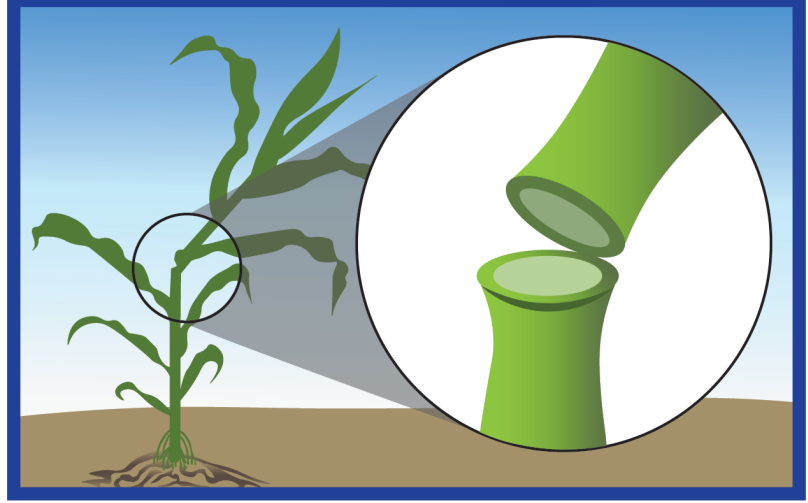

**D - Maize brazier bucking (late-season)**

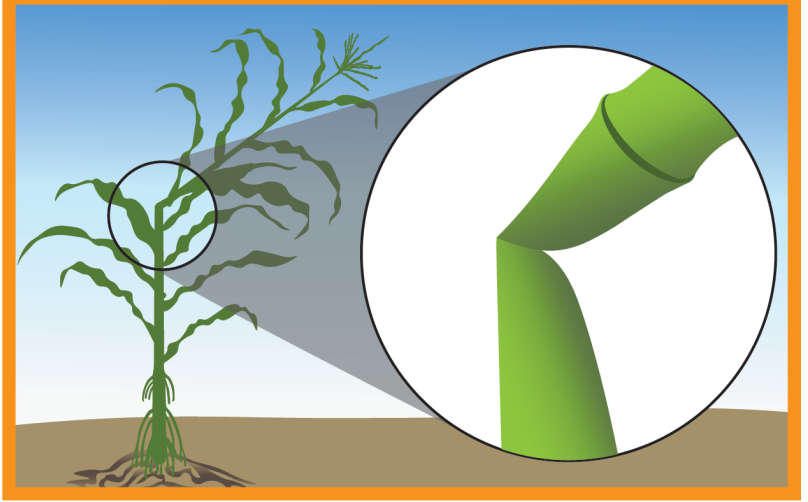

**A - Uprooting**

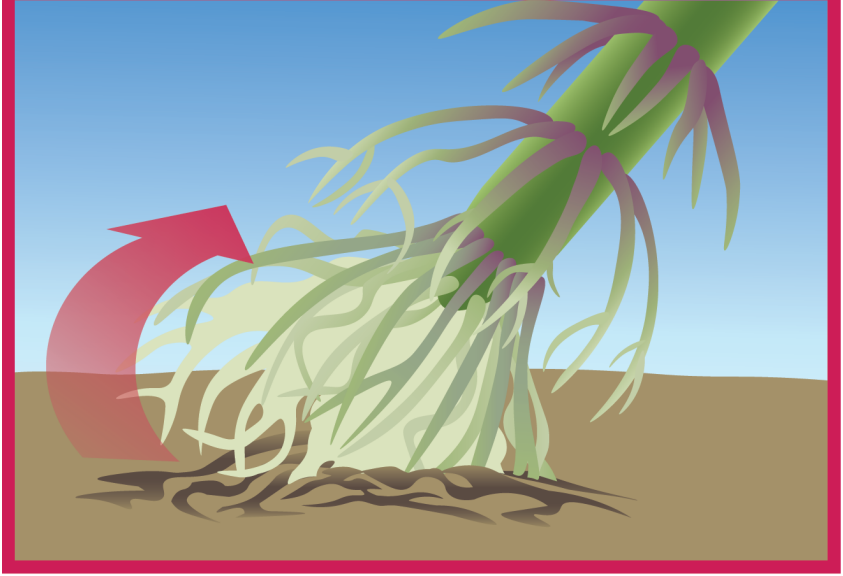

**B - Breakage**

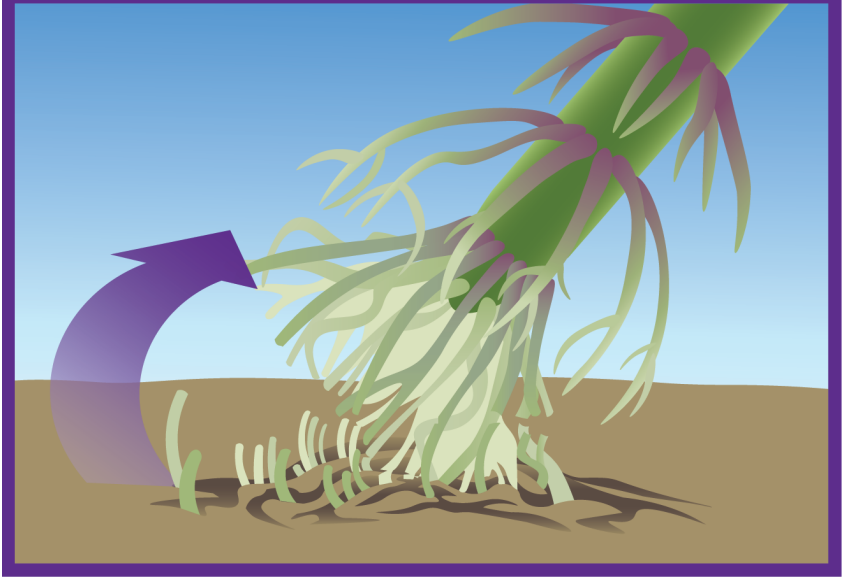

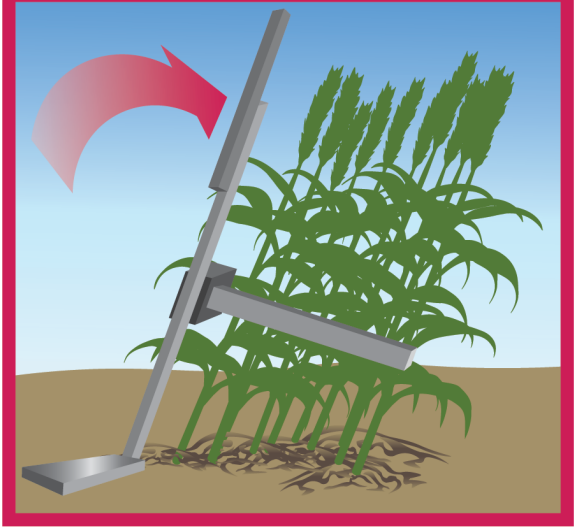

**A - Berry's Device**

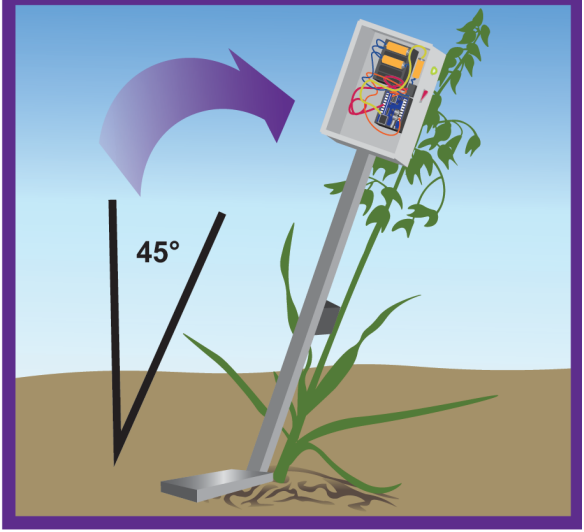

**B - The Stalker**

45°

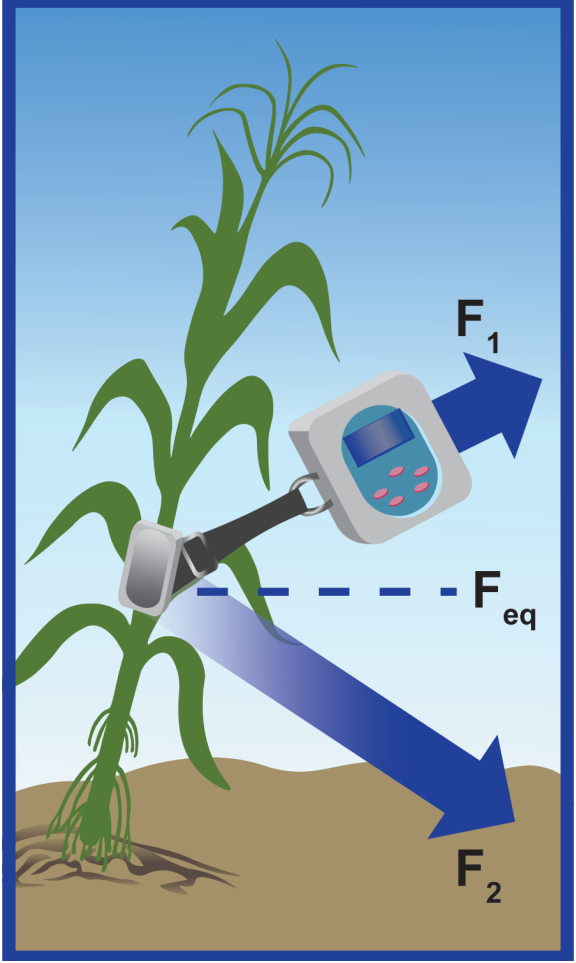

**C - Guo's Device**

$F_1$

$F_{eq}$

$F_2$

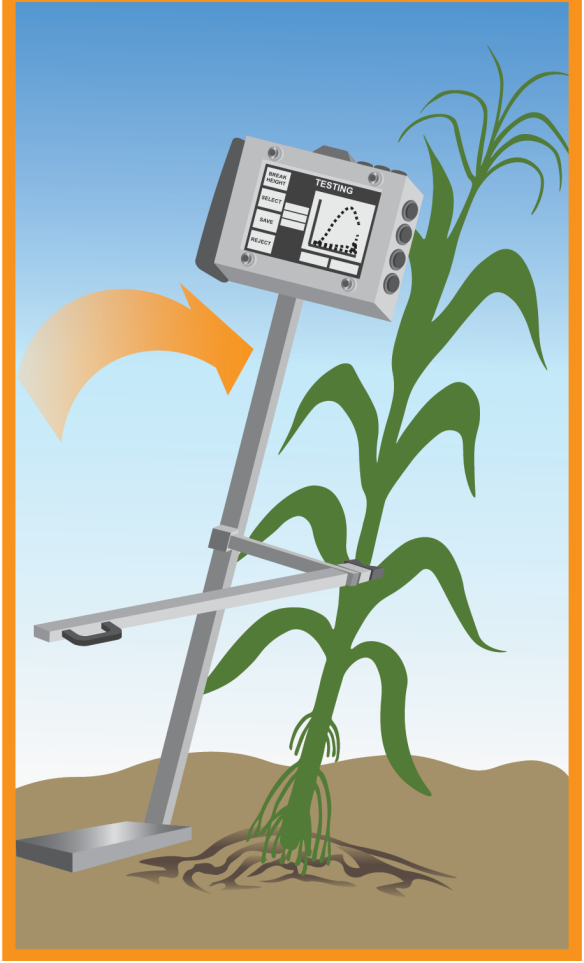

**D - DARLING**

TESTING

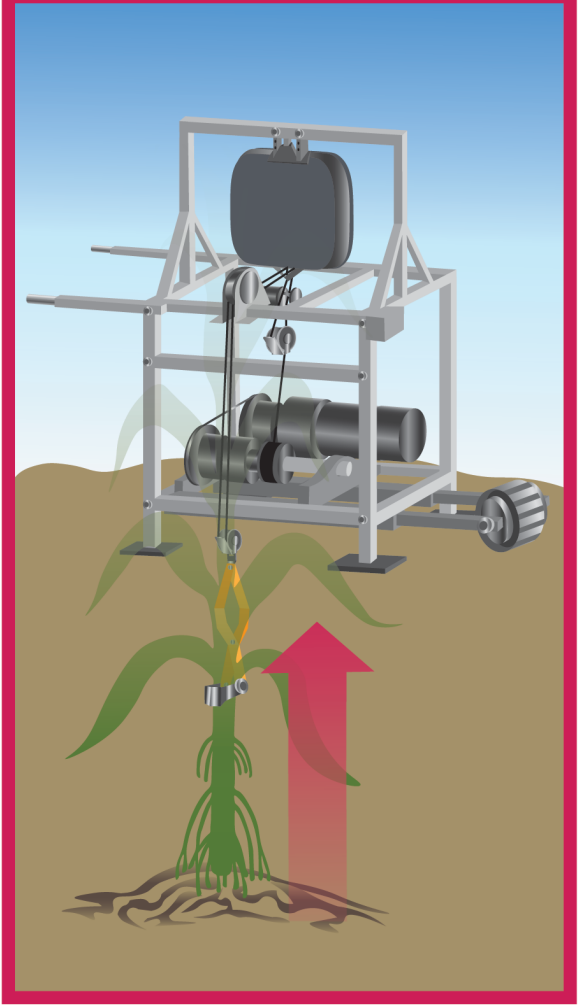

**A - Dourleijn's Device**

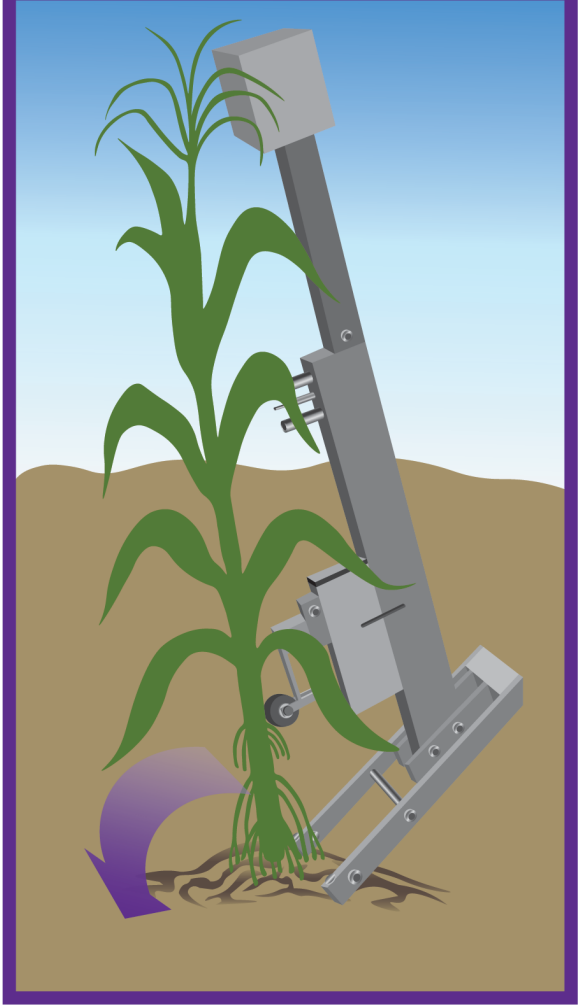

**B - Fouéré's Device**

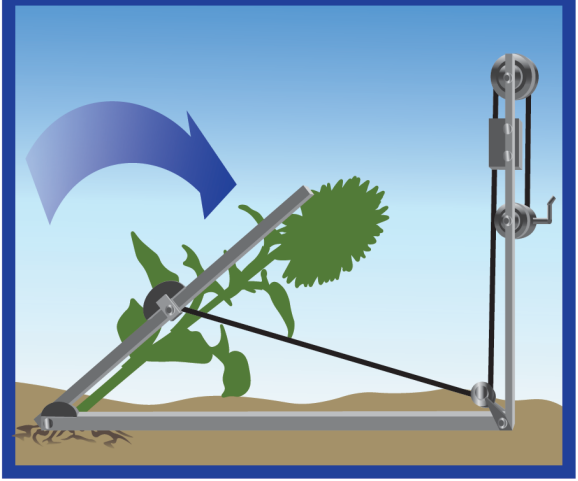

**C - Sposaro's Device**

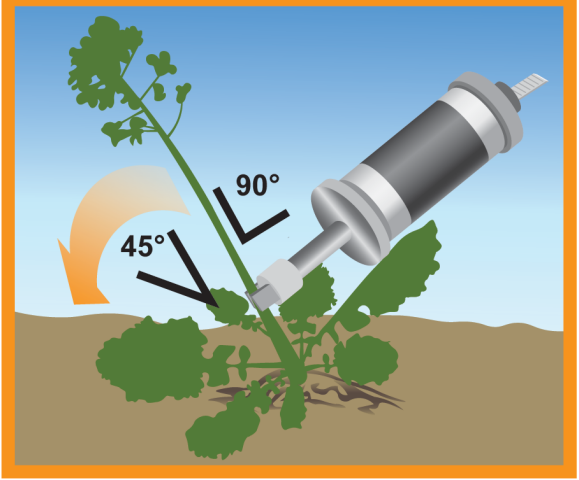

**D - Prostrate Testing Device**

90°

45°